\documentclass[conference]{IEEEtran}
\IEEEoverridecommandlockouts
\usepackage{cite}

\usepackage{amsmath,amssymb,amsfonts}
\usepackage{algorithmic}
\usepackage{graphicx}
\usepackage{textcomp}
\usepackage{xcolor}
\def\BibTeX{{\rm B\kern-.05em{\sc i\kern-.025em b}\kern-.08em
    T\kern-.1667em\lower.7ex\hbox{E}\kern-.125emX}}

\begin{document}

\title{Prescribed-Time Observer Is Naturally Robust Against Disturbances and Uncertainties}

\author{
\IEEEauthorblockN{ABEDOU Abdelhadi}
\IEEEauthorblockA{
Paris-Saclay University
}
\and
\IEEEauthorblockN{MAMECHE Omar}
\IEEEauthorblockA{
Paris-Saclay University
}
}

\maketitle

\begin{abstract}
This paper addresses the robustness of a prescribed-time observer for a class of nonlinear systems in the presence of disturbances and unmodeled dynamics. It is  proven and demonstrated through simulations that the proposed observer completely rejects the effects of arbitrarily large bounded disturbances and unmodeled dynamics, enabling accurate estimation of both the states and the disturbances. Furthermore, a comparison with the standard high-gain observer is provided to highlight the superiority of the prescribed-time observer in reducing the peaking phenomenon and improving estimation accuracy.

\end{abstract}


\section{Introduction}

State estimation for non-linear systems is a fundamental and challenging problem in control, monitoring, and fault diagnosis, as the full state of the system is often not directly measurable but crucial for effective control design. To address this, observer-based methods have been extensively studied and applied. Classic observer designs, such as the extended Kalman observer \cite{kalman1960new}, the Luenberger observer \cite{1099826}, the high-gain observer \cite{4694705,10054587,9655232}, the sliding mode observer \cite{spurgeon2008sliding}, and the adaptive observer \cite{989154,ZHUANG2023126865,GHANI20235685,article2}, provide powerful tools for estimating the states of the system under different assumptions and conditions. These observers vary in their convergence properties and robustness, and selecting an appropriate design depends on the system dynamics and application requirements. Despite the variety of approaches, challenges remain in achieving fast, reliable, and robust state estimation in non-linear systems, especially when practical constraints such as disturbances and model uncertainties are present.

\vspace{0.2cm}

Observers can be broadly categorized based on their convergence behavior into asymptotic and finite-time convergent types. In recent years, finite-time state estimation has gained considerable interest due to the strict time response demands in many practical control applications. Several finite-time observer designs have been developed to meet these requirements. Among them, sliding mode observers are well-regarded for their inherent robustness against system uncertainties and external disturbances \cite{LEVANT1998379,ANGULO20132489,5893919}. Homogeneity-based observers, on the other hand, leverage the structural properties of the system to ensure finite-time convergence \cite{5399809,LOPEZRAMIREZ201852,6137934,MENARD2017438,ANDRIEU2009422}. In addition, algebraic approaches, such as those that utilize modulating functions, have also been proposed to achieve fast state estimation \cite{7320661,GHAFFOUR20204196,BELKHATIR2018451,doi:10.1049/iet-cta.2018.5313}. Despite these advances, a key challenge remains: The convergence time of finite-time observers is often strongly influenced by the magnitude of the initial estimation error, which can lead to slower transient responses when the initial error is large.

\vspace{0.2cm}

To overcome this, prescribed-time observers have been developed in the literature. These observers guarantee that the estimation error converges to zero within a fixed time that is defined a priori, independently of the initial conditions. Initial work focused on linear systems using time-varying gains and coordinate transformations to achieve fixed-time convergence \cite{8600393}. More recent extensions to non-linear triangular systems employ state transformations with time-scaling functions and exploit their properties to ensure global asymptotic stability with convergence within the prescribed time \cite{CHEN2020104640,article,chitour2019stabilizationperturbedchainintegrators,9666587,10549778}.

\vspace{0.2cm}

Although the design and convergence properties of prescribed-time observers have been well studied, their robustness against disturbances and model uncertainties has not been established. This robustness is crucial for practical applications where systems are subject to external perturbations and modeling errors. Motivated by this gap, our paper rigorously proves that prescribed-time observers are naturally robust against disturbances and uncertainties. We demonstrate that the intrinsic structure of the prescribed-time observer completely rejects the influence of external perturbations and modeling errors, ensuring estimation error convergence within the prescribed time despite these factors. This natural robustness is a key advantage over traditional observer designs, making prescribed-time observers especially well suited for real-world problems affected by disturbances and unmodeled dynamics.

\vspace{0.2cm}

The second section of this paper, titled \textbf{Preliminaries and Problem Formulation}, introduces the necessary mathematical background and formalizes the state estimation problem addressed. The third section, \textbf{Robustness Analysis of the Prescribed-Time Observer}, presents our main  contribution by proving the natural robustness properties of the prescribed-time observer against disturbances and uncertainties. Finally, the last section, \textbf{Numerical Examples}, compares the performance of the prescribed-time observer with the classical high-gain observer, focusing on transient response and peaking phenomenon, and examine how both observers handle disturbances and unmodeled dynamics. In this section, we also discuss how we can extand the prescribed-time observer for exact simultaneous state and disturbance estimation.

\section{Preliminaries and Problem Formulation}
\noindent In this section, we introduce some useful preliminaries and formulate the estimation problem.

\subsection{Preliminaries}
\noindent We will recall some definitions and lemma, which are necessary for the mathematical developments given in the next sections.

\textit{\textbf{Definition 1}} \cite{CHEN2020104640}: The system $\dot{x}=f(x,t)$ is said to be globally convergent to zero in any prescribed-finite time $T$, if for any initial state $x(0)\in\mathbb{R}^{n}$, the system states are well defined on $t\in[0,T)$ and satisfy
\[\lim_{t\to T}\|x(t)\|=0.\]

\textit{\textbf{Lemma 1}} \cite{6473848}: Suppose the matrix $D\in\mathbb{R}^{n\times n}$ is defined as $D=\text{diag}\{1,\ldots,n\}$ and $P\in\mathbb{R}^{n\times n}$ is a positive definite matrix. Then, a positive constant $\lambda$ exist and satisfy \begin{center}
$PD + DP \geq \lambda I$.
\end{center}

\noindent The following function is introduced for the estimation design $\mu(t,T):[0,T)\to[1,\infty )$ as \cite{8600393}
\[\mu(t,T):=\frac{T}{T-t}.\]
This function is monotonically increasing having the property that $\mu(t,T)$ tends to infinity as $t\to T$ where $T$ is predefined time.

\subsection{Problem Formulation}
\noindent We consider the class of nonlinear systems described by

\begin{equation}
\left\{
\begin{array}{l}
\dot{x}_1 = x_2 + f_1(x_1, u) \\
\vdots \\
\dot{x}_{n-1} = x_n + f_{n-1}(x_1, \ldots, x_{n-1}, u) \\
\dot{x}_n = f_n(x_1, \ldots, x_n, u) + d(t) \\
y = x_1
\end{array}
\right.
\label{eq:system}
\end{equation}

\noindent where $x(t) = [x_1,x_2,\dots,x_n]^T\in\mathbb{R}^{n}$ is the state vector of the system and $y(t)\in\mathbb{R}$ is the measured output.

\vspace{0.2cm}

\noindent We can rewrite the system $(1)$ in the form: 
\begin{equation}
\left\{
\begin{array}{l}
\dot{x}_1 = x_2 + f_1(x_1, u) \\
\vdots \\
\dot{x}_{n-1} = x_n + f_{n-1}(x_1, \ldots, x_{n-1}, u) \\
\dot{x}_n = f_0(x_1, \ldots, x_n, u) + \sigma(t) \\
y = x_1
\end{array}
\right.
\label{eq:system}
\end{equation}

\noindent Where $\sigma(t) = f_n(x_1, \ldots, x_n, u) - f_0(x_1, \ldots, x_n, u) + d(t)$, and $f_0(\cdot)$ is a nominal model of $f_n(\cdot)$.

\vspace{0.2cm} 

\noindent The prescribed-time observer will be designed under the following assumptions:

\vspace{0.2cm}

\textit{\textbf{Assumption 1}:}  The disturbances $d(t)$ is assumed unknown but bounded with unknown bounds: 
\[
\exists \,\bar{d} \in \mathbb{R}^+ , \; \|d(t)\| \leq \bar{d} 
\]

\vspace{0.1cm}

\textit{\textbf{Assumption 2}:}  The functions $f_{i}:\mathbb{R}^{i}\longrightarrow\mathbb{R}$, $i=0, 1,...,n$ satisfy the local Lipschitz property formulated under the following form
\[||f_{i}(x_{1},\ldots,x_{i},u)-f_{i}(\hat{x}_{1},\ldots,\hat{x}_{i},u)||\leqslant \gamma_{f_{i}} \sum_{k=1}^{i} ||x_k-\hat{x}_k||\]
where $\gamma_{f_{i}}$ is the Lipschitz constant of the function $f_i$.

\vspace{0.2cm}

\textit{\textbf{Assumption 3}:}  The states $x(t)$ and their estimates $\hat{x}(t)$ are defined in a compact set.

\vspace{0.2cm}

\noindent Using the above assumptions, it is clear that $\sigma(t)$ is bounded: 

\[
\exists \,\bar{\sigma} \in \mathbb{R}^+ , \; \|\sigma(t)\| \leq \bar{\sigma} 
\]

\vspace{0.2cm}

\textit{\textbf{Remark 1}:} Note that $\gamma_{f_1},\gamma_{f_2},\dots,\gamma_{f_n}$ and $\bar{d}$ are arbitrary large bounded constants.

\vspace{0.2cm}

\noindent The prescribed-time observer for system (1) is given by : 
\begin{equation}
\left\{
\begin{array}{l}
\dot{\hat{x}}_1 = \hat{x}_2 + \hat{f}_1 + L_1 \cdot \mu^{1+m}(t) \cdot (y - \hat{y}) \\
\vdots \\
\dot{\hat{x}}_{n-1} = \hat{x}_n + \hat{f}_{n-1} + L_{n-1} \cdot \mu^{(n-1)(1+m)}(t) \cdot (y - \hat{y}) \\
\dot{\hat{x}}_n = \hat{f}_0 + L_n \cdot \mu^{n(1+m)}(t) \cdot (y - \hat{y}) \\
\hat{y} = \hat{x}_1
\end{array}
\right.
\label{eq:observer}
\end{equation}

\noindent Where $\hat{f}_i = f_i(\hat{x}_1,\dots, \hat{x}_i,u)$, $ \forall i=0,...,n$, $[L_1,\dots,L_n]$ are the observer's gains, and $m$ is a design parameter such that $m > 0$.

\vspace{0.2cm}

\noindent The objective of this paper is to prove that the prescribed-time observer (3) can completely rejects the effects of arbitrarily large bounded disturbances and unmodeled dynamics on the estimation error.

\section{Robustness Analysis of the Prescribed-Time Observer}

\noindent This section focuses on the analysis of estimation error dynamics and the robustness properties of the proposed observer.

\subsection{Estimation Error Dynamics}

\noindent Let the estimation error be defined as $e(t) = x(t) - \hat{x}(t)$. Its dynamics are given by:
\begin{equation}
\left\{
\begin{array}{l}
\dot{e}_1 = e_2 + {f}_1 - \hat{f}_1
 - L_1 \cdot \mu^{1+m}(t) \cdot e_1 \\
\vdots \\
\dot{{e}}_{n-1} = {e}_n + {f}_{n-1} - \hat{f}_{n-1} - L_{n-1} \cdot \mu^{(n-1)(1+m)}(t) \cdot e_1 \\
\dot{{e}}_n = {f}_0 - \hat{f}_0 + \sigma(t) - L_n \cdot \mu^{n(1+m)}(t) \cdot e_1 \\
\hat{y} = \hat{x}_1
\end{array}
\right.
\label{eq:observer}
\end{equation}

\noindent In order to prove the prescribed-time convergence of the observer in (3), and to establish its robustness to disturbances and uncertainties by showing that their effect on the estimation error can be completely removed, the following state transformation is used:
\[z(t)=\Gamma{e}(t).\]

\noindent where $\Gamma$ is a scaling matrix defined as:
\[\Gamma=diag\{1/\mu^{1+m},\ldots,1/\mu^{n(1+m)}\}\] 

\vspace{0.2cm}

\noindent The dynamics of the transformed error is given as follows
\[\dot{z}(t)=\mu^{1+m}(A-LC)z(t)-(1+m)\frac{\dot{\mu}}{\mu}Dz(t)+\Gamma\Delta f + \Gamma B \sigma(t)\]
where $D$ is a diagonal matrix defined as:
\[
D={diag}\{1,2,\ldots,n\}
\]

\noindent
The matrices \( A \), \( B \), and \( C \) are given as follows: \vspace{0.1cm}
\[
A = \begin{bmatrix}
0 & 1 & 0 & \cdots & 0 \\
0 & 0 & 1 & \cdots & 0 \\
\vdots & \vdots & \vdots & \ddots & \vdots \\
0 & 0 & 0 & \cdots & 1 \\
0 & 0 & 0 & \cdots & 0
\end{bmatrix} \quad
B = \begin{bmatrix}
0 \\
0 \\
\vdots \\
0 \\
1
\end{bmatrix} \quad
C = \begin{bmatrix}
1 & 0 & \cdots & 0
\end{bmatrix}
\]

\vspace{0.1cm}

\subsection{Robustness Analysis}
This part is devoted to the main theorem, which provides synthesis conditions guaranteeing prescribed-time stability of the observer error.

\vspace{0.2cm}

\textit{\textbf{Theorem :}}
Let the gain matrix $L$ be designed such that $A-LC$ is hurwitz, then the observer error is asymptotically stable at defined time $T>0$, and the observer completely rejects the effects of arbitrarily large bounded disturbances and unmodeled dynamics. 

\vspace{0.2cm}

\textit{\textbf{Proof :}}
Consider the following Lyapunov function candidate
\[V(z(t))=z(t)^{T}Pz(t),\quad P>0.\]
The derivative of V along the trajectories (12) can be calculated as
\begin{align*}
\frac{\partial V}{\partial t} &= z^TPz + z^TP\dot{z} \\
&= \mu^{1+m}z^T\left[ \left( A - LC \right)^TP + P(A - LC) \right]z \\
&- \left( 1 + m \right)\frac{\dot{\mu}}{\mu}z^T\left[ D^TP + PD \right]z + 2z^TP\Gamma \big[\Delta f + B \sigma(t) \big] 
\end{align*}

\noindent There exist positive constants \( \lambda_1 \) and \( \lambda_2 \) such that:
\begin{align*}
\left( A - LC \right)^TP + P(A - LC) &\leq -\lambda_1 I_n, \\
D^TP + PD &\geq \lambda_2 I_n.
\end{align*}

\noindent Hence the derivative of $V(t)$ becomes
\[ \frac{\partial V}{\partial t} \leq -\lambda_1\mu^{1+m}z^Tz - \lambda_2(1 + m)\frac{\dot{\mu}}{\mu}z^Tz + 2z^TP\Gamma\left[\Delta f + B \sigma(t)\right] \] 

\noindent Using the fact that $f$ is Lipschitz, we have: 
\begin{align*}
\frac{\|\Delta f_i(x, \hat{x}, u)\|}{\mu^{i(1+m)}} 
&\leq \frac{1}{\mu^{i(1+m)}} 
\Big\| f_i(x_1, \ldots, x_i, u) - {f}_i(\hat{x}_1, \ldots, \hat{x}_i, u) \Big\| \\
&\leq \frac{\gamma_{f_i}}{\mu^{i(1+m)}} \cdot \sum_{k=1}^{i} ||x_k-\hat{x}_k|| \\
&\leq \frac{\gamma_{f_i}}{\mu^{i(1+m)}} \cdot \sum_{k=1}^{i} \mu^{k(1+m)} \cdot ||z_k-\hat{z}_k|| \\
&\leq \gamma_{f_i} \sum_{k=1}^{i} \mu^{(k-i)(1+m)} \cdot ||z_k-\hat{z}_k||
\end{align*}
\noindent For all $i = 1, \ldots, n$, this implies the existence of $\bar{\gamma}_f$, which is independent of time for all $\mu(t) \geq 1$ and $i \geq k$, such that:
\begin{align*}
\|\Gamma \Delta f \| & \leq \bar{\gamma}_f \cdot \|z\| 
\end{align*}
\noindent Thus,
\begin{align*}
2z^TP\left[\Gamma \Delta f\right] & \leq 2\|z\| \cdot \|P\| \cdot \|\Gamma \Delta f\| \\ 
&\leq 2 \cdot \bar{\gamma}_f \cdot \lambda_{\max}(P) \cdot \|z\|^2 \\
&\leq a \cdot \|z\|^2
\end{align*}

\noindent Where $a = 2 \cdot \bar{\gamma}_f \cdot \lambda_{\max}(P)$

\vspace{0.2cm}

\noindent Using the fact that $\sigma(t)$ is bounded, we have
\begin{align*}
2z^TP\left[\Gamma B \sigma(t) \right] & \leq 2\|z\| \cdot \|P\| \cdot \|\Gamma B \| \cdot \|\sigma(t)\| \\ 
& \leq 2\|z\| \cdot \lambda_{max}(P) \cdot \|\Gamma B \| \cdot \|\sigma(t)\| \\ 
& \leq  \frac{ 2 \cdot \bar{\sigma} \cdot \lambda_{max}(P) }{\mu^{n(1+m)}} \cdot \|z\| \\ 
& \leq  \frac{b}{\mu^{n(1+m)}} \cdot \|z\| 
\end{align*}

\noindent Where $b = 2 \cdot \bar{\sigma} \cdot \lambda_{max}(P) $

\vspace{0.2cm}

\noindent Knowing that:
\[\frac{\dot{\mu}}{\mu} = \frac{1}{T}\mu\]

\noindent We get the following:
\begin{align*}
- \lambda_2(1 + m)\frac{\dot{\mu}}{\mu}z^Tz & = - \frac{\lambda_2}{T} (1 + m) \cdot \mu \cdot \|z\|^2 \leq 0 
\end{align*}

\noindent We obtain the following inequality:
\begin{align*}
\frac{\partial V}{\partial t} & \leq -\lambda_1\mu^{1+m}\|z\|^2 + a\|z\|^2 + \frac{b}{\mu^{n(1+m)}} \cdot \|z\|
\end{align*}

\noindent Which can be rewritten as:
\begin{align}
\frac{\partial V}{\partial t} 
& \leq -\lambda_1\mu^{1+m}\|z\|^2 
+ a\|z\|^2 
+ \frac{b \cdot \mu^{1+m}}{\mu^{(n+1)(1+m)}} \cdot \|z\|
\label{eq:V_derivative}
\end{align}

\noindent We distinguish two cases based on the norm of \( z(t) \):

\vspace{0.3cm}

\noindent \textit{\textbf{1.}} \(\|z(t)\| < \dfrac{2b}{\lambda_1 \cdot \mu^{(n+1)(1+m)}}\) \textit{\textbf{(Inside the Ball)}} \\

\noindent \textit{\textbf{2.}} \(\|z(t)\| \geq \dfrac{2b}{\lambda_1 \cdot \mu^{(n+1)(1+m)}}\) \textit{\textbf{(Outside the Ball)}}

\vspace{0.3cm}

\noindent We begin by considering the simpler case, namely, when the $\|z\|$ is inside the ball.

\vspace{0.2cm}

\noindent \textit{\textbf{1. Inside The Ball:}}

\vspace{0.1cm}

\noindent We have: 
\begin{flalign*}
\|z(t)\| & \leq \dfrac{2b}{\lambda_1 \cdot \mu^{(n+1)(1+m)}} \to 0, \quad as \quad t \to T 
\end{flalign*}

\noindent Moreover,
\begin{flalign*}
\|e(t)\| & \leq \mu^{n(1+m)} \cdot \|z(t)\| && \\
& \leq \mu^{n(1+m)} \cdot \frac{2b}{\lambda_1 \cdot \mu^{(n+1)(1+m)}} && \\
& \leq \frac{2b}{\lambda_1 \cdot \mu^{(1+m)}} \to 0, \quad as \quad t \to T  &&
\end{flalign*}

\vspace{0.2cm}

\noindent We get:
\[
\lim_{t \to T} \|e(t)\| = 0
\]

\vspace{0.1cm}

\noindent Therefore, if we are inside the ball, it shrinks to zero over time, and the estimation error \( e(t) \) converges to zero in prescribed-time $T$.

\vspace{0.3cm}

\noindent We now move to the second case (outside the ball). Since we have already shown that, inside the ball, the ball shrinks to zero and \( e(t) \to 0 \) as \( t \to T \), it remains to prove that, outside the ball, the trajectory of $z(t)$ enters the ball in finite time $T$ and that \( e(t) \) also tends to zero as \( t \to T \).

\vspace{0.2cm}

\noindent \textit{\textbf{2. Outside The Ball:}}

\vspace{0.1cm}

\noindent We have: 
\begin{flalign*}
\|z(t)\| & \geq \dfrac{2b}{\lambda_1 \cdot \mu^{(n+1)(1+m)}}
\end{flalign*}

\noindent Substituting into equation~\eqref{eq:V_derivative}, we obtain:
\[
\frac{\partial V}{\partial t}  \leq \left( -\frac{\lambda_1}{2} \cdot \mu^{1+m} + a \right)\cdot V \quad \quad \]

\noindent Because $\mu(t)$ is monotonically increasing, a time $0<t_{1}^*<T$ exists such that:
\[ \frac{\lambda_1}{4} \cdot \mu^{1+m} \geq a \]

\noindent For $t \in [0, t_{1}^*]$, one can directly have:
\[
\frac{\partial V}{\partial t}  \leq a\cdot V\]
\vspace{-0.5cm}
\begin{flalign*}
V(t) &\leq V(0) \cdot \exp  \left( a \cdot t \right) 
\end{flalign*}
\vspace{-0.5cm}
\begin{align}
\|z(t)\| 
&\leq \sqrt{\frac{V(0)}{\lambda_{\min}(P)}} 
\cdot \exp \left( \frac{a}{2} \cdot t \right)
\label{eq:z_bound}
\end{align}

\vspace{-0.2cm}

\begin{align}
\|e(t)\| 
&\leq \sqrt{\frac{V(0)}{\lambda_{\min}(P)}} 
\cdot \mu^{n(1+m)} \cdot \exp \left( \frac{a}{2} \cdot t \right)
\label{eq:e_bound}
\end{align}

\noindent This means that the finite-time escape phenomenon will not occur for $t \in [0,t_{1}^*]$ as the right-hand side of equation~\eqref{eq:z_bound} and ~\eqref{eq:e_bound} is well defined.

\vspace{0.2cm}

\noindent For $t \in [t_{1}^*,T)$, we obtain:
\[
\frac{\partial V}{\partial t}  \leq -\frac{\lambda_1}{4} \cdot \mu^{1+m} \cdot V \quad \quad \]
\vspace{-0.5cm}
\begin{flalign*}
V(t) &\leq V(t_{1}^*) \cdot \exp\left( -\frac{\lambda_1 T}{4m} \left( \mu_t^{m} - \mu_{t_*}^{m} \right) \right)
\end{flalign*}

\noindent Therefore,
\begin{flalign*}
\|z(t)\| &\leq \sqrt{\frac{V(t_{1}^*)}{\lambda_{\min}(P)}} 
\cdot \exp\left( -\frac{\lambda_1 T}{8m} \left( \mu_t^{m} - \mu_{t_*}^{m} \right) \right)
\end{flalign*}

\noindent
Since \( \mu(t) \) is monotonically increasing, there exists a time 
\( t_2^* \in (t_1^*, T) \) such that  
$$\|z(t_2^*)\| < \frac{2b}{\lambda_1 \, \mu^{(n+1)(1+m)}}$$

This implies that, even if the trajectory starts outside "the ball", 
there exists a time \( t_2^* < T \) at which it enters the ball.  
As the ball shrinks to the origin at time \( T \), and all trajectories 
inside the ball converge to \( 0 \), we conclude that
\[
\lim_{t \to T} \|z(t)\| = 0,
\qquad \text{and therefore} \qquad
\lim_{t \to T} \|e(t)\| = 0.
\]

\vspace{0.2cm}

\noindent According to Definition 1, we conclude that the observer error \( e(t) \) is globally asymptotically stable at the prescribed time \( T \), even in the presence of uncertainties in the model and disturbances, highlighting the inherent robustness of the observer. Moreover, we have shown that the finite-time escape phenomenon does not occur. As a result, the prescribed-time observer does not diverge, despite its gains tending to infinity as \( t \to T \). This ends the proof.

\section{Numerical Examples}
To show the performance of the prescribed-time observer design in the presence of disturbances and uncertainties, we present two numerical examples in this section. The simulations are carried out using MATLAB.

\subsection{Example 1}

\noindent The aim of this example is to demonstrate the effectiveness of the proposed prescribed-time observer in estimating the states of a nonlinear system in triangular form, compared to a high-gain observer, in the presence of model uncertainties and external disturbances.

\vspace{0.2cm}

\noindent The considered nonlinear system is defined as follows:
\begin{equation}
\left\{
\begin{aligned}
\dot{x}_1 &= x_2 - l_1 \sin(x_1) \\
\dot{x}_2 &= -x_1 - l_2 x_2^3 + u + d(t) \\
y &= x_1
\nonumber
\end{aligned}
\right.
\end{equation}

\noindent The values of the system parameters are set to $l_1 = 1$, $l_2 = 0.02$. The control input is $u(t) = sin(0.35t)$. $d(t)$ denotes an external disturbance $ d(t) = 5sin(2t)$.

\vspace{0.2cm}

\noindent The prescribed-time observer can be constructed
as follows:
\begin{equation}
\left\{
\begin{aligned}
\dot{\hat{x}}_1 &= \hat{x}_2 - l_1 \sin(\hat{x}_1) + L_1 \cdot  \mu^{1+m} \cdot (x_1-\hat{x}_1) \\
\dot{\hat{x}}_2 &= 0 + L_2 \cdot \mu^{2(1+m)} \cdot (x_1-\hat{x}_1)
\nonumber
\end{aligned}
\right.
\end{equation}

\noindent The numerical parameters for the \textbf{prescribed-time observer} are selected as:
{\setlength{\abovedisplayskip}{3pt}%
\setlength{\belowdisplayskip}{3pt}%
\[
L_1 = 3, \quad L_2 = 2, \quad T = 0.5, \quad m = 0.1
\]%
}
\noindent The high-gain observer can be constructed as follows:
{\setlength{\abovedisplayskip}{3pt}%
\setlength{\belowdisplayskip}{3pt}%
\begin{equation}
\left\{
\begin{aligned}
\dot{\hat{x}}_1 &= \hat{x}_2 - l_1 \sin(\hat{x}_1) + \frac{\alpha_1}{\varepsilon} \cdot (x_1-\hat{x}_1) \\
\dot{\hat{x}}_2 &= 0 + \frac{\alpha_2}{\varepsilon} \cdot (x_1-\hat{x}_1)
\nonumber
\end{aligned}
\right.
\end{equation}
}
\noindent The high-gain observer parameters are chosen such that, after convergence and settling, the estimation error norm remains small, effectively attenuating the effect of uncertainties and disturbances. These parameters are set as follows:
{\setlength{\abovedisplayskip}{3pt}%
\setlength{\belowdisplayskip}{3pt}%
\[
\alpha_1 = 3, \quad \alpha_2 = 2, \quad \varepsilon = 0.01
\]
}

\noindent We compare the performance of both observers in terms of their ability to rejects disturbances and compensate for uncertainties, with particular attention to the \textit{peaking phenomenon} exhibited during the transient phase.

\vspace{0.2cm}

\begin{figure}[!ht]
    \centering
    \includegraphics[width=0.45\textwidth]{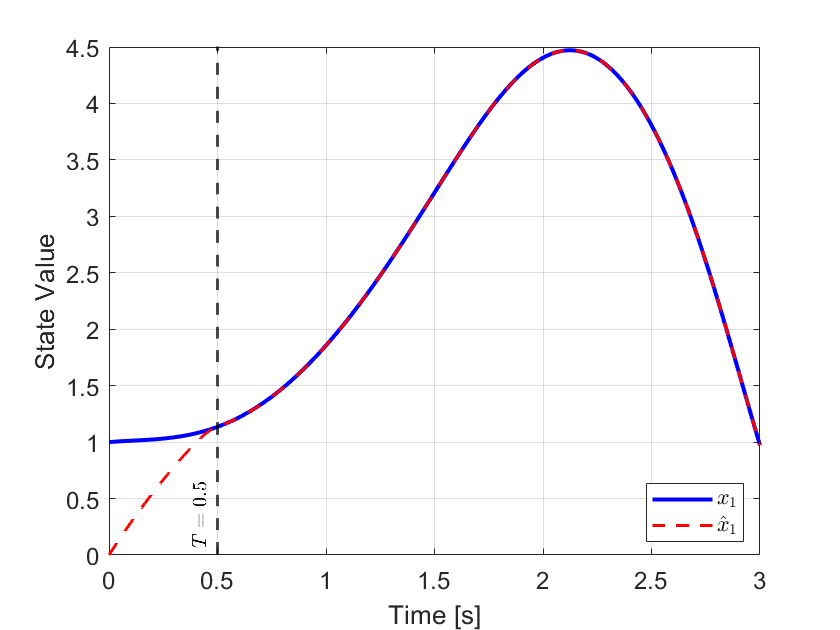}
    \caption{System state $x_1$ and its estimate $\hat{x}_1$ using the prescribed-time observer.}
    \label{fig:sim1_x1}
\end{figure}

\begin{figure}[!ht]
    \centering
    \includegraphics[width=0.45\textwidth]{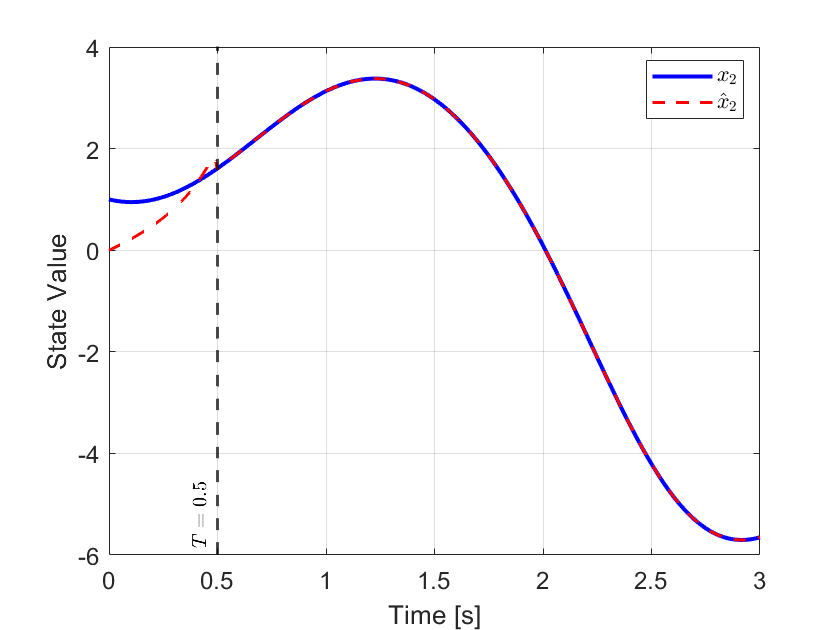}
    \caption{System state $x_2$ and its estimate $\hat{x}_2$ using the prescribed-time observer.}
    \label{fig:sim1_x2}
\end{figure}

\begin{figure}[!ht]
    \centering
    \includegraphics[width=0.45\textwidth]{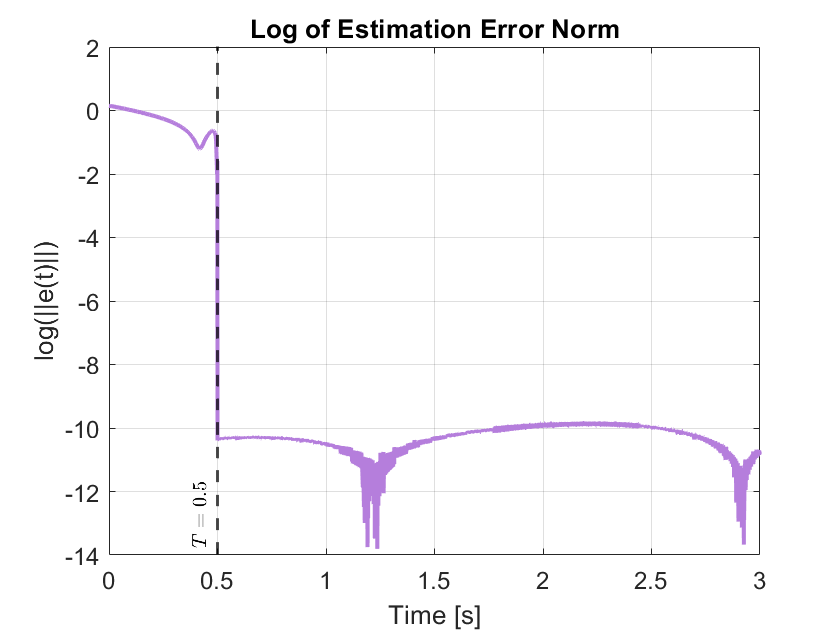}
    \caption{Norm of the estimation error $\|x - \hat{x}\|$ using the prescribed-time observer.}
    \label{fig:sim1_loge}
\end{figure}

\noindent Figures~\ref{fig:sim1_x1}, \ref{fig:sim1_x2}, and \ref{fig:sim1_loge} illustrate the performance of the prescribed-time observer.

\vspace{0.2cm}

\noindent Figures~\ref{fig:sim1_x1} and \ref{fig:sim1_x2} show the system states $x_1$, $x_2$ and their corresponding estimates $\hat{x}_1$, $\hat{x}_2$, respectively. Figure~\ref{fig:sim1_loge} presents the norm of the estimation error $\|x - \hat{x}\|$.

\vspace{0.2cm}

\noindent It is important to note that the parameter $\mu$ was saturated during the simulations to prevent numerical instabilities. In this particular case, its value was limited to $10^{10}$.

\begin{figure}[!ht]
    \centering
    \includegraphics[width=0.45\textwidth]{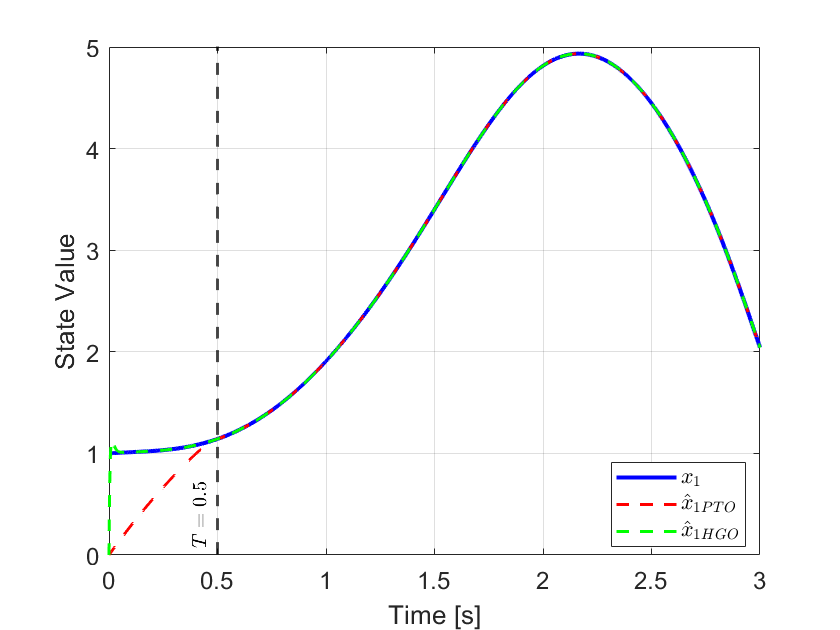}
    \caption{System state $x_1$ and its estimate $\hat{x}_1$ comparison.}
    \label{fig:sim3_x1}
\end{figure}

\begin{figure}[!ht]
    \centering
    \includegraphics[width=0.45\textwidth]{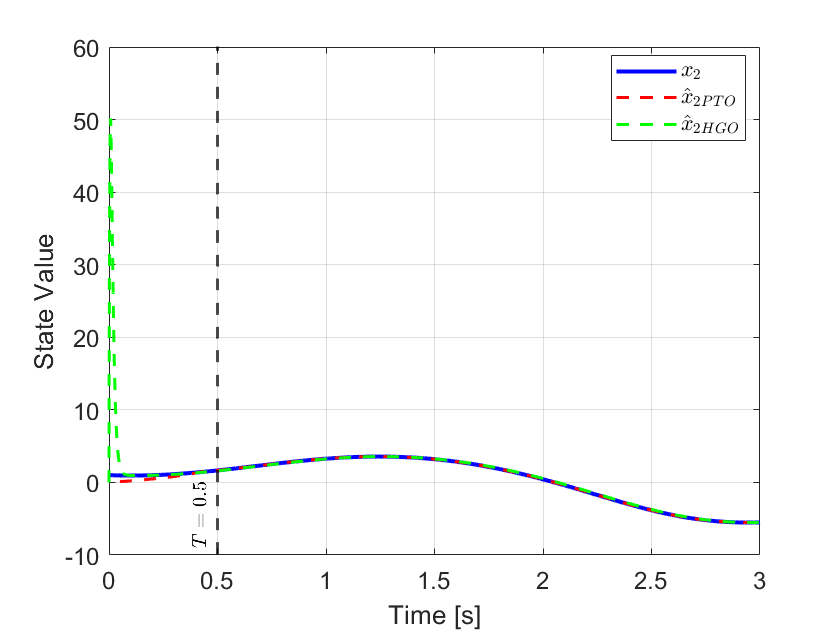}
    \caption{System state $x_2$ and its estimate $\hat{x}_2$ comparison.}
    \label{fig:sim3_x2}
\end{figure}

\begin{figure}[!ht]
    \centering
    \includegraphics[width=0.45\textwidth]{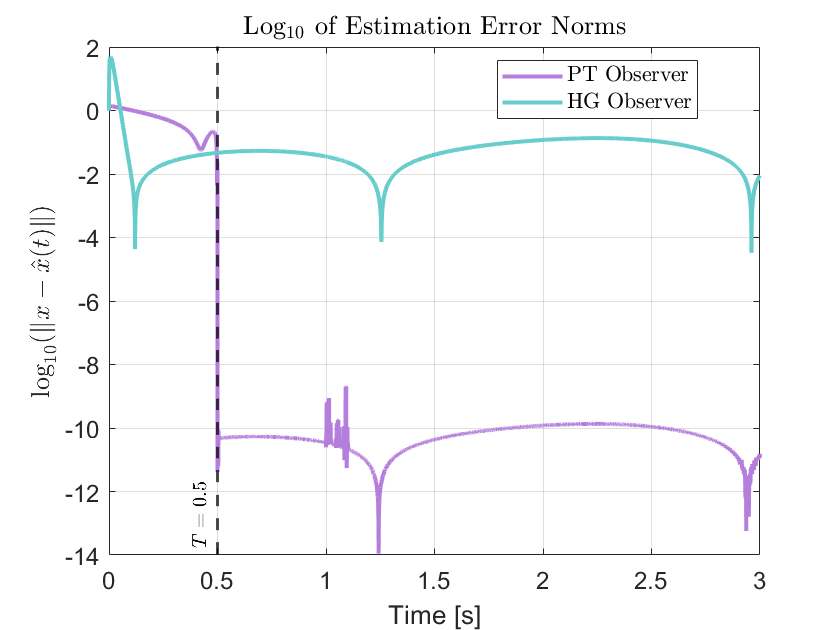}
    \caption{Estimation error norm $\|x - \hat{x}\|$ comparison.}
    \label{fig:sim3_loge}
\end{figure}

\noindent Figures~\ref{fig:sim3_x1}, \ref{fig:sim3_x2}, and \ref{fig:sim3_loge} compare the performance of the prescribed-time observer and the high-gain observer.

\noindent Figures~\ref{fig:sim3_x1} and \ref{fig:sim3_x2} depict the system states \( x_1 \), \( x_2 \) alongside their estimates \( \hat{x}_1 \), \( \hat{x}_2 \), respectively. Figure~\ref{fig:sim3_loge} shows the norm of the estimation error \( \|x - \hat{x}\| \).

\noindent From the results, we observe that for \( \varepsilon = 0.01 \), the peaking phenomenon is significantly more pronounced in the high-gain observer case. Additionally, the steady-state error is less effectively attenuated with the high-gain approach. This comparison demonstrates the superior performance of the prescribed-time observer in both transient and steady-state behavior.

\subsection{Example 2}

\noindent This example shows how a prescribed-time observer estimates disturbances by augmenting the system with a disturbance state while preserving its triangular structure.

\vspace{0.2cm}

\noindent The considered nonlinear system is the same as in\textit{ Example 1}. The prescribed-time observer is constructed as follows:
\begin{equation}
\left\{
\begin{aligned}
\dot{\hat{x}}_1 &= \hat{x}_2 - l_1 \sin(\hat{x}_1) + L_1 \cdot  \mu^{1+m} \cdot (x_1-\hat{x}_1) \\
\dot{\hat{x}}_2 &= -\hat{x}_1 - l_2 \hat{x}_{2}^3 + u + \hat{d} + L_2 \cdot \mu^{2(1+m)} \cdot (x_1-\hat{x}_1) \\
\dot{\hat{d}} &= 0 + L_3 \cdot \mu^{3(1+m)} \cdot (x_1-\hat{x}_1)
\nonumber
\end{aligned}
\right.
\end{equation}
\noindent The \textbf{prescribed-time observer} parameters are:
{\setlength{\abovedisplayskip}{3pt}%
\setlength{\belowdisplayskip}{3pt}%
\[
L_1 = 6, \quad L_2 = 11, \quad L_3 = 6, \quad T = 1, \quad m = 0.1
\]
}

\begin{figure}[!ht]
    \centering
    \includegraphics[width=0.45\textwidth]{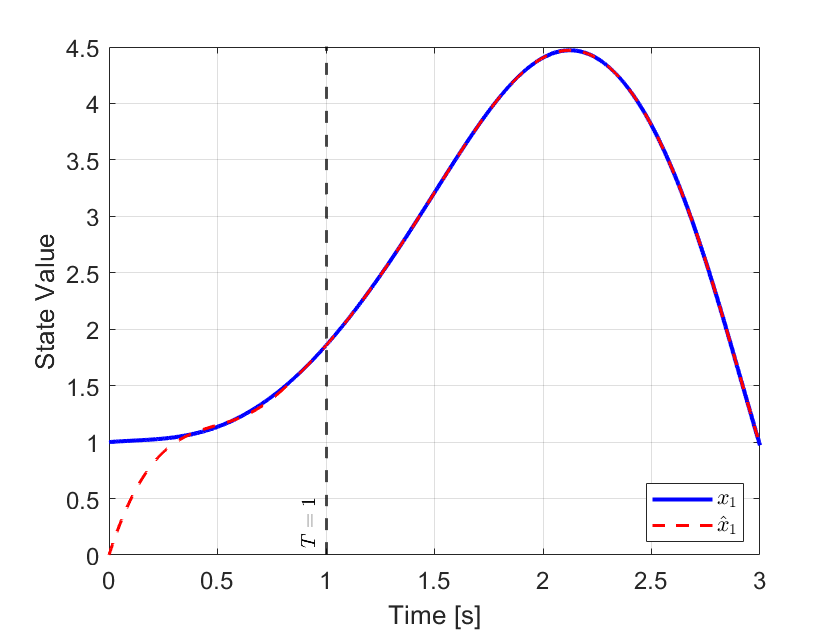}
    \caption{$x_1$ and its estimate $\hat{x}_1$ using the extended prescribed-time observer.}
    \label{fig:sim2_x1}
\end{figure}

\begin{figure}[!ht]
    \centering
    \includegraphics[width=0.45\textwidth]{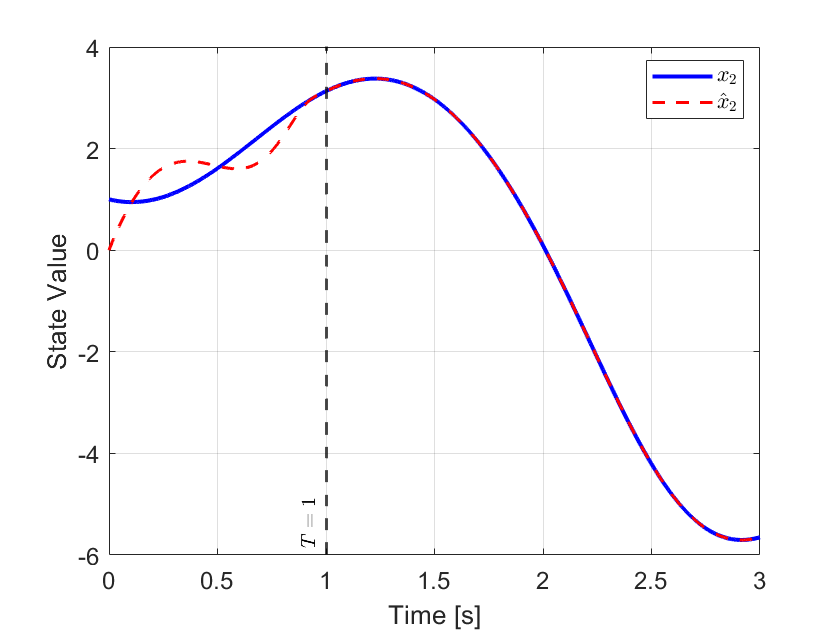}
    \caption{$x_2$ and its estimate $\hat{x}_2$ using the extended prescribed-time observer.}
    \label{fig:sim2_x2}
\end{figure}

\begin{figure}[!ht]
    \centering
    \includegraphics[width=0.45\textwidth]{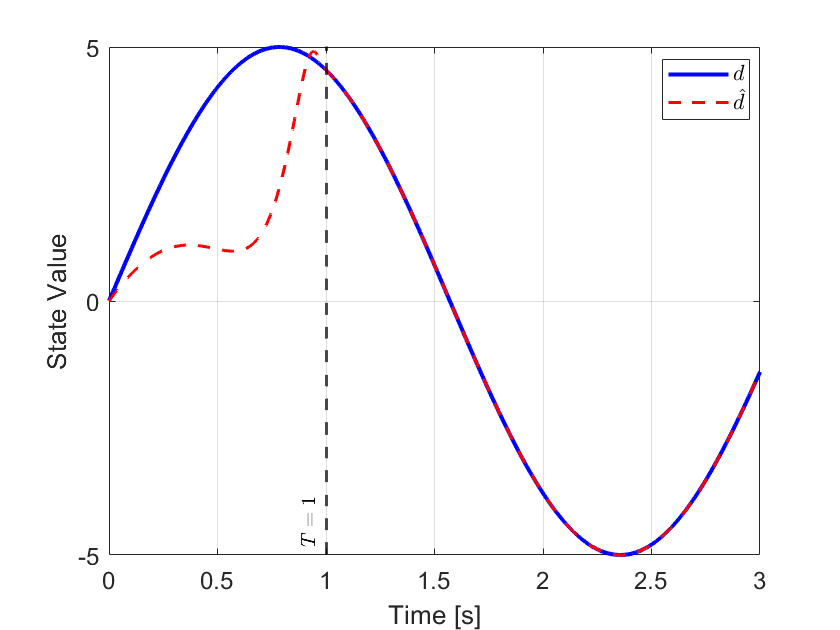}
    \caption{The disturbance $d$ and its estimate $\hat{d}$ using the extended prescribed-time observer.}
    \label{fig:sim2_d}
\end{figure}

\noindent Figures~\ref{fig:sim2_x1}, \ref{fig:sim2_x2}, and \ref{fig:sim2_d} illustrate the performance of the prescribed-time observer by showing the system states $x_1$, $x_2$, and the disturbance $d$, along with their corresponding estimates $\hat{x}_1$, $\hat{x}_2$, and $\hat{d}$, respectively.

\newpage
\section{CONCLUSION}
In this work, the robustness of a prescribed-time observer for a class of nonlinear systems has been rigorously analyzed in the presence of bounded disturbances and modeling uncertainties. It has been  proven and confirmed through numerical examples that the observer completely rejects the effects of such uncertainties and disturbances, ensuring accurate estimation of both the states of the system and the disturbance within a predefined convergence time $T$ selected a priori. Furthermore, a comparison with the standard high-gain observer highlights the superior performance of the prescribed-time observer in mitigating the peaking phenomenon and enhancing estimation accuracy.

\bibliographystyle{ieeetr}
\bibliography{references}

\end{document}